# Determination of Tritium Content in $D_2$ Gas and Heavy Water

Sandeep Puri[a]*, Noah D'Amico[a], Ian Jones[a], Sonny Alaniz[b], Andrew Gillespie[a], Cuikun Lin[a], R. V. Duncan[a]

[a] *Center for Emerging Energy Sciences, Department of Physics and Astronomy, Texas Tech University, Lubbock, Texas, USA*

[b] *Department of Physics and Astronomy, Texas Tech University, Lubbock, Texas, USA*

*** Corresponding Authors:** sanpuri@ttu.edu

**Keywords:** Tritium, Nuclear Fusion, Deuterium Gas, Quantulus

**Abstract**

A simple experimental method for the determination of tritium concentration in natural hydrogen and deuterium gas down to the femtomole level is described. Palladium and platinum catalysts were used to react hydrogen and oxygen, creating water that could be analyzed in a Revvity Quantulus GCT 6220. This analysis produced sub-femtomole detection limits of tritium present in hydrogen gas or heavy water. A contamination level of ~1 femtomole/SL was found in the $D_2$ gas sample that was analyzed. A contamination level of ~0.025 femtomole/SL, which is below the noise level of the Quantulus, was found in the standard hydrogen gas sample that was analyzed. The choice of palladium or platinum catalyst did not significantly affect the resulting tritium measurement. By showing that common $D_2$ samples can be contaminated with trace amounts of tritium orders of magnitude above natural abundance, this test demonstrates that a baseline tritium measurement for deuterium samples is highly important for determining the validity of results in nuclear fusion yield experiments.

## 1.0 Introduction

Tritium ($^3H$) is a radioisotope (half-life ~12.32 yr) that is naturally produced in Earth's atmosphere when high energy cosmic neutrons collide with $^{14}N$, creating tritium and carbon-12. The tritium is quickly oxidized, forming $^3HHO$ and becoming part of the global water cycle with a natural relative abundance of $10^{-18}$. Nuclear bomb testing between 1945-1963 led to a 3-4 order of magnitude increase in the abundance of tritium, though tritium levels have slowly declined back to baseline levels since the 1963 Partial Nuclear Test Ban Treaty.[1–3] The natural abundance of tritium is too low for isotope separation of water to be an efficient source of the radioisotope, so the primary method for tritium production occurs within nuclear reactors, where lithium-6 is split by a neutron into tritium and an alpha particle.

Tritium plays a critical role in nuclear fusion experiments and environmental monitoring. Establishing proper hydrogen isotope ratios for D-T fusion in large gas samples is a relatively simple task that does not require extreme levels of measurement resolution. Therefore, a wide range of established practices exist for this measurement. Deuterium ($^2H$, or D), a stable isotope of hydrogen, is also valuable for nuclear experiments. Heavy water ($D_2O$) is used as a neutron moderator and coolant in nuclear reactors due to its low atomic mass and low neutron absorption rate. Previously, we analyzed tritium content of heavy water samples from various venders. The samples contain tritium on the order of ~1 femtomole/gram of $D_2O$ or a relative abundance near $10^{-14}$ (10 parts-per-quadrillion), and the results of these measurements are displayed in **Table 1**.

| Sample | **Mass [g]** | **CPM/g** | **Post Calibration [DPM/g]** | **Tritium Concentration [fmole/g]** |
|---|---|---|---|---|
| Acros Organic Lot: A0435170 100% D2O | 4.0896 | 6.92 | 38.09 | 0.60 |
| Acros Organic Lot: A0434911 99.8% D2O | 4.6325 | 13.61 | 74.85 | 1.18 |
| TCI Lot: 4JTCA-PD 99.8% D | 4.8812 | 13.10 | 72.02 | 1.14 |
| Aldrich Lot: MKBT5294V 99.9% D | 6.3237 | 14.32 | 78.74 | 1.24 |
| ThermoScientific Lot: Y08J026 99.8% D | 1.0565 | 11.95 | 65.7 | 1.03 |

**Table 1:** Mass, counts per gram measured, adjusted calculated activity, and tritium concentration per gram of various $D_2O$ samples. Sample provider and batch number are listed.

It is this neutron-exposed heavy water that is often later separated into $D_2$ gas and collected in dewars for various lab and industrial applications. Therefore, gaseous deuterium fuel for D-D fusion must be analyzed for tritium content with high resolution to rule out the possibility of tritium contamination leading to false nuclear fusion diagnostics, a factor that is often overlooked in Low Energy Nuclear Reactions (LENR). D-D fusion has two product pathways which are split with 50/50 probabilities:

$$D + D \rightarrow {}^3He + {}^1n + 3.27\ MeV \quad (1)$$

$$D + D \rightarrow {}^3H + {}^1p + 4.03\ MeV. \quad (2)$$ [4]

Similar diagnostics are also needed for heavy water ($D_2O$), which can be a product of fusion experiments or current commercial nuclear reactors.[5] The existing methods for these measurements include Beta-ray Induced X-ray Spectrometry (BIXS), Mass Spectrometry (MS), gas ionization chambers, which all present significant limitations.[1,6–10]

In this paper, we describe a simple experimental method for the determination of tritium concentration in protium or deuterium gas down to the femtomole level. The hydrogen (any isotope) gas sample is combined in a 1:4 ratio with oxygen, using the excess oxygen as a carrier gas. This mixed gas flows over a heated palladium or platinum recombiner, after which it is condensed and collected in two cold traps.[11] The samples can then be analyzed for tritium content in the Revvity Quantulus GCT 6220, an ultra-low-level liquid scintillation spectrometer designed for measuring extremely small activities of beta- and alpha-emitting radionuclides with high precision.[12,13] It uses advanced background-reduction technologies, including passive and active shielding, to achieve exceptionally low detection limits. The instrument provides excellent energy resolution and is widely used in environmental, radiological, and nuclear research where high-sensitivity measurements are required. This process can be applied to gas samples before and after fusion experiments to test for contamination and detect sub-femtomole levels of tritium caused by low fusion reaction rates that are commonly expected in LENR.

Both platinum and palladium catalysts were used to test recent experimental results from other groups regarding enhanced deuterium fusion rates in electrolysis using palladium.[14,15] No difference in tritium concentration was found after electrolysis between the catalysts, so we believe no measurable fusion occurred with the palladium catalyst.

**2.0 Experiment**

Flow controllers were used to mix 50 mL/min of either natural hydrogen gas or deuterium gas with 200 mL/min oxygen. Carried by the excess oxygen, the mixture flowed into a 3/8" copper tube filled with either palladium (Alfa Aesar, 5% on 3mm alumina pellets) or platinum (Thermo Scientific, 0.5% on 3mm alumina pellets). The tube was twisted into a coil, heated to 105 °C using heating tape and insulated with ceramic wool. After the gases reacted, water vapor was condensed in an ice bath cold trap. The remaining vapor was collected in a liquid nitrogen trap. A diagram of the experimental setup is shown in **Figure 1**.

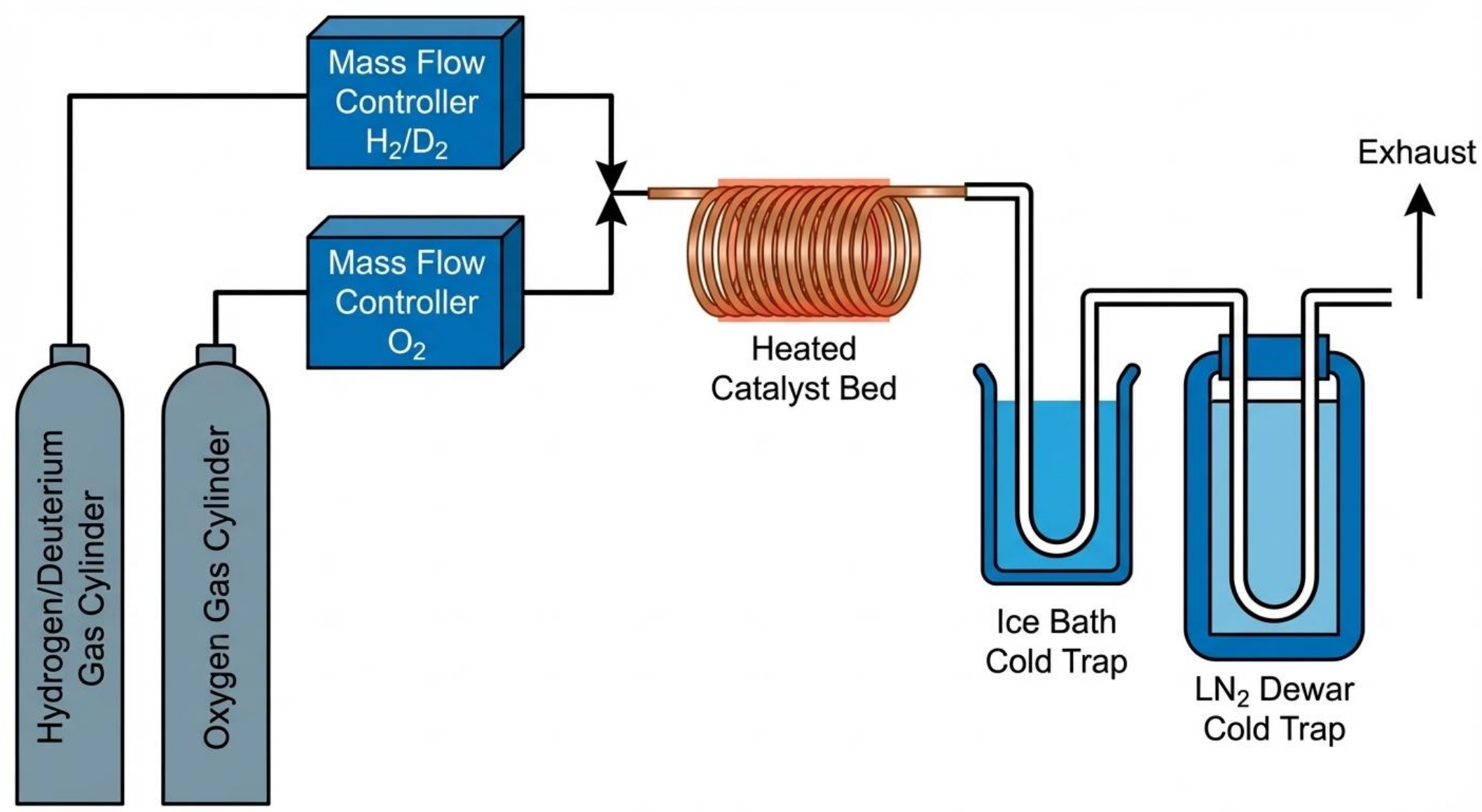

**Figure 1:** Diagram of experimental setup for reaction of $H_2/D_2$ and $O_2$ using Pt or Pd catalysts.

The resulting water (5-10 mL) was collected, transferred into vials, and combined with 10 mL Ultima Gold liquid scintillating cocktail.[16] If needed, additional HPLC water was added to increase the total volume to 20 mL. This mixture was shaken until it was no longer cloudy. The vials were then placed in a Revvity Quantulus GCT 6220[12,13] that had been calibrated with various known tritium samples beforehand. The samples were analyzed, and the data was recorded in terms of DPM (disintegrations per minute). This measure can easily be converted into mass or moles, as is done in the Results section.

**3.0 Results and Discussion**

The activity of each water sample in CPM (counts per minute), DPM and femtomoles/SL of hydrogen gas is displayed in **Table 2**. A sample of known tritium concentration was used to determine the relationship between measured CPM and actual activity, DPM, from the Quantulus instrument (1 CPM = 5.5 DPM). Previous calibration of the Quantulus has shown that samples with known activity below 3 CPM cannot be accurately distinguished from samples of a known activity of 0 CPM, leading to a detection limit of about 0.04 femtomoles/SL. Note that the activity of the $H_2$ samples is just below the accepted detection limit of the Quantulus, so we are unable to conclude its true concentration, except for stating that it is below 0.04 fmole/SL.

| Sample | Water Mass [g] | SL of $H_2$ or $D_2$ [L] | Total CPM | Adjusted DPM | Tritium Concentration [fmole/SL] |
|---|---|---|---|---|---|
| **$H_2$ on Pd** | 6.61 | 8.22 | 2.4 | 13.2 | 0.027 |
| **$H_2$ on Pt** | 7.84 | 9.75 | 2.36 | 13 | 0.022 |
| **$D_2$ on Pd** | 9.09 | 11.31 | 121.45 | 668 | 0.917 |
| **$D_2$ on Pt** | 8.24 | 10.25 | 112.8 | 620.4 | 0.94 |

**Table 2:** Mass, volume of $H_2$ or $D_2$ gas at STP consumed, total counts measured, adjusted calculated activity, and tritium concentration per standard liter of $H_2$ and $D_2$ gas samples with both palladium and platinum catalysts.

It is clear that stock deuterium gas is contaminated with small amounts of tritium above the natural abundance concentration on earth. The natural abundance of tritium on earth is about $10^{-18}$, while our data shows that there is a relative abundance of $\sim 10^{-14}$ (10 parts-per-quadrillion) in deuterium gas samples. A relative tritium abundance of $<10^{-15}$ was found in hydrogen gas samples, though this is on the order of the noise level of the Quantulus, so the true contamination may be lower. These data prove that scientists exploring LENR must look for concentrations of tritium well above 1 part per $10^{14}$ as evidence of D-D fusion, not just above 1 part per $10^{18}$ as could be assumed from the natural abundance.

## 4.0 Conclusion

The results show that deuterium gas may contain ~1 femtomole/SL tritium contamination. There is no difference in recombination performance when using a palladium or platinum catalyst. There may also be trace amounts (~0.025 femtomole/SL) of tritium contamination in hydrogen gas, though this figure is around the noise level of the measurement system, so more precise analysis would be needed for a definitive result. The similar levels of tritium concentration in $D_2$ gas samples and heavy water samples lead us to believe that the origin of the tritium contamination is the nuclear reactors from which the heavy water is sourced, though this is not confirmed in this study; this information is proprietary and protected by the sourcing company.

## 5.0 Conflict of Interest

The authors declare that the research was conducted in the absence of any commercial or financial relationships that could be construed as a potential conflict of interest.

## 6.0 Author Contributions

All authors contributed to the conception and design of the research reported here. ND wrote the first draft of the manuscript. All authors contributed to manuscript revision, read, and approved the submitted version.

## 7.0 Funding

This work was supported by the Department of Energy award No. DE-AR0001736.